\documentclass{aa}
\usepackage{graphicx}
\usepackage{url}

\usepackage{txfonts}
\usepackage[breaklinks,colorlinks,citecolor=blue]{hyperref}
\usepackage{natbib,twoopt}
\newcommand{\GG}[1]{}

\bibpunct{(}{)}{;}{a}{}{,} 
\makeatletter
\newcommandtwoopt{\citeads}[3][][]{\href{http://adsabs.harvard.edu/abs/#3}%
{\def\hyper@linkstart##1##2{}%
\let\hyper@linkend\@empty\citealp[#1][#2]{#3}}}
\newcommandtwoopt{\citepads}[3][][]{\href{http://adsabs.harvard.edu/abs/#3}%
{\def\hyper@linkstart##1##2{}%
\let\hyper@linkend\@empty\citep[#1][#2]{#3}}}
\newcommandtwoopt{\citetads}[3][][]{\href{http://adsabs.harvard.edu/abs/#3}%
{\def\hyper@linkstart##1##2{}%
\let\hyper@linkend\@empty\citet[#1][#2]{#3}}}
\newcommandtwoopt{\citeyearads}[3][][]%
{\href{http://adsabs.harvard.edu/abs/#3}
{\def\hyper@linkstart##1##2{}%
\let\hyper@linkend\@empty\citeyear[#1][#2]{#3}}}
\makeatother

\newcommand{\sqdeg}{deg$^2$}

\def \arcsec      {\text{$^{\prime\prime}$}}
\def \mjybeam     {mJy\,beam$^{-1}$}

\begin{document}

   \title{The `spectral index-flux density relation' for extragalactic radio sources selected at metre and decametre wavelengths}

   \titlerunning{Spectral index-flux density relations}

\author {Pratik Dabhade\inst{1,2}\thanks{E-mail: pdabhade@iac.es}
\and Gopal-Krishna\inst{3}
}
  
\institute{$^{1}$Instituto de Astrof\' isica de Canarias, Calle V\' ia L\'actea, s/n, E-38205, La Laguna, Tenerife, Spain\\
$^{2}$Universidad de La Laguna (ULL), Departamento de Astrofisica,
E-38206, Tenerife, Spain\\
$^{3}$UM-DAE Centre for Excellence in Basic Sciences (CEBS), Vidyanagari, Mumbai - 400098, India\\ 
}

 \date{\today} 
 
 \abstract
{We use the recent releases of sensitive VLA/LOFAR large-area surveys at 340 MHz and 54 MHz, in conjunction with the 1.4 GHz NVSS, to accurately determine the `spectral index - flux density relation' ($\alpha$ - S) for extragalactic radio sources selected at metre and decametre wavelengths, the latter for the first time. This newly determined $\alpha$ - S$_{\rm 340~MHz}$ relation shows a progressive flattening of
$\alpha_{\rm median}$ towards lower flux densities, starting from its steepest value (peak)occurring near S$_{\rm 340~MHz}$ $\sim$\, 1-2 Jy. This resolves the controversy extant in the literature since the 1980s. The $\alpha$ - S$_{\rm 54~MHz}$ relation, too, shows a spectral index flattening with decreasing flux density which, however, is significantly milder and the relation is less sharply peaked than that found at 340 MHz. A possible reason for the difference could be that the 54 MHz sample has a  distinctly stronger/conspicuous presence ( at $\sim$\,20\% level) of very steep spectrum sources having $\alpha_{54}^{1400} <$ -1.3, most of which are probably associated with clusters of galaxies. }


\maketitle

\section{Introduction} \label{sec:intro}
Statistical variation of radio spectral index ($\alpha$ ; S~$\propto\nu^{\alpha}$) of extragalactic radio sources with  monochromatic radio luminosity (P) was inferred in several studies, starting from those conducted more than five decades ago \citep[][]{Heeschen1960,kpw69,Bridle1972,Veron1972,MacLeod1972,GK88,Blundell1999}. A statistically more significant correlation was reported in \citet{Laing1980} and modelled in terms of multiple shocks occurring within the hot spots \citep{GKW1990}. In an alternative claim, the correlation of spectral index with redshift, $z$ was posited to be more basic \citep[][]{Katgert1980,Macklin1982}.  Choosing between these two alternatives has remained difficult mainly because in a flux-limited sample, P and $z$ are strongly correlated due to Malmquist bias \citep[e.g.][]{Singal1993,Blundell1999}. Another complicating factor is the statistical trend for the spectra of individual sources to be curved, becoming progressively steeper between metre and centimetre wavelengths \citep[][]{Laing1980,GK88,Blundell1999,Konar2006,Jamrozy2008,McKean2016}. An attempt to resolve this degeneracy was made using two flux-limited samples selected at metre wavelengths, for which radio spectrum of each source could be determined in the rest-frame; it was thus inferred that the correlation of $\alpha$ with P is likely to be primary (\citealt{GK88} ; also, \citealt{Blundell1999}). These above correlations involving $\alpha$ have a bearing on a number of important issues, such as the strong tendency (i) of high-$z$ radio galaxies to exhibit a very steep radio spectrum ($\alpha$ $<$ - 1.1 ) at decimetre wavelengths \citep{Tielens1979,Blumenthal1979,GKS81,DeBreuck2000}, and (ii) for the very steep spectra to remain straight at least up to $\sim$\,10 GHz \citep{Mangalam1995,Klamer2006}.

Another well known statistical dependence of $\alpha$ is on radio flux density (S). The variation of $\alpha$ with flux density can be used to constrain models of cosmological evolution of extragalactic radio sources \citep[][]{Dagkesamanskii1970,Laing1980,Condon1984,Kulkarni1990,Calistro2017}. For samples of sources selected at metre wavelengths, statistical dependence of $\alpha$ on S was first pointed out almost 5 decades ago by \citet{Murdoch1976}. Using a complete sample of 133 sources above a limiting flux density of 0.97 Jy at 408 MHz, he found the mean value of $\alpha_{408}^{2695}$ to be -0.911$\pm$0.014, which is significantly steeper than the corresponding value of -0.816$\pm$0.018 found for the All-sky Catalogue of \citet{Robertson1973}, which is a complete sample of sources stronger than 10 Jy at 408 MHz. For metre-wavelength selected sources, this demonstrated a 4.2$\sigma$ level systematic spectral steepening between the strong and intermediate flux densities. This led to the obvious question: does the steepening of $\alpha$ continue to (say, an order-of-magnitude) lower flux levels, or does it plateau out, or even reverses? This question was first investigated in a set of two papers on `spectral index - flux density relation' using metre-wavelength samples (\citealt{GS82}:\texttt{GS82} ; \citealt{SG84}: \texttt{SG84}), spanning a wide range in flux density from S$_{408}$ $\sim$\,25 Jy to $\sim$\,0.1 Jy which is known to be dominated by distant extragalactic radio sources \citep{Condon1989}. In these two papers, particularly the latter one in which the authors used their more accurate flux measurements of a complete sample of 212 sources, made with the 100-metre Effelsberg radio telescope, it was shown that $\alpha_{\rm median}$ steepens from -0.815$\pm$0.02 at about S$_{408}$ $\sim$\,15 Jy to -0.90$\pm$0.02 at intermediate flux levels (S$_{408}$ $\sim$\, 1 - 2 Jy), and thereafter flattens back 
to $\sim$\, -0.78$\pm$0.03 by S$_{408}$ $\sim$\,0.2 Jy (mirroring the differential source counts of radio sources selected at metre wavelengths). While this trend was confirmed by \citet{Windhorst1990}, \citet{Kapahi1986} and later \citet{Kulkarni1990} found the flattening of $\alpha_{\rm median}$ between S$_{408}$ $\sim$\,2 Jy and $\sim$\,0.2 Jy to be roughly an order-of-magnitude smaller, hence barely significant \citep[see, also,][]{Zhang2003}. In yet another study, based on 1.5 GHz VLA imaging 
of 1103 sources from the Bologna B3 sample at 408 MHz \citep{Ficarra1985}, \citet{Vigotti1989} found that the steepening of $\alpha$ (mean) continues all the way down to S$_{408}$ $\sim$\,0.1 Jy. Thus, whilst almost all the studies from 1975-1990 were consistent in finding for metre-wavelength samples a systematic steepening of $\alpha$ from high to intermediate (S$_{408}$ $\sim$\,1 - 2 Jy) flux density levels, contradictory trends were claimed about the run of $\alpha$  towards a further order-of-magnitude lower flux densities. During the past 5 years, this issue has been revisited by \citet{Tiwari2019} and \citet{deGasperin2018}, using the 147 MHz survey TIFR GMRT SKY SURVEY Alternate Data Release 1 \citep[TGSSADR1;][]{intema-tgss-17}, by combining this database with the 1.4 GHz NRAO VLA Sky Survey catalogue \citep[NVSS;][]{nvss}. Both these independent analyses confirm a significant spectral flattening occurring between the intermediate and lower flux densities, consistent with the results of \texttt{GS82} and \texttt{SG84}\footnote{The sequence of data points defining the $\alpha_{147}^{1400}$-S$_{147}$ relation has a large gap near S$_{147}$ $\sim$\,2 Jy \citep{deGasperin2018} where the relation is expected to peak (i.e., steepest $\alpha$). Nonetheless, the trend indicated by their data points seems consistent with the expected peaking of $\alpha$.}.  One may, however, recall that the spectral curvature of sources often becomes very significant at frequencies below $\sim$\,200 MHz, as underscored, e.g., in \citet{Laing1980} and this raises the possibility of bias getting introduced in the $\alpha$-S relation determined in the above mentioned two recent studies which are based on TGSS-ADR1 survey made at 147 MHz. This gave us the impetus to verify the $\alpha$-S relation using a similarly large metre-wavelength survey carried out at a somewhat higher frequency, namely the recent VLITE\footnote{\href{https://vlite.nrao.edu/}{VLA Low-band Ionosphere and Transient Experiment}} Commensal Sky Survey Epoch 1 (`VCSS1') at 340 MHz, see \citet{Polisensky2016} and \citet{Clarke2016SPIE} for details. In addition, the recent release of the LOFAR LBA Sky Survey preliminary data \citep[LoLSS-PDR][]{Gasperin2021} survey at 54~MHz has enabled us to extend the $\alpha$ - S relation to decametre wavelengths. The higher-frequency data have been taken from the 1.4 GHz NRAO VLA Sky Survey
catalogue \citep[NVSS;][]{nvss}\footnote{We use the S~$\propto\nu^{\alpha}$ convention for calculating $\alpha_{340}^{1400}$ $\&$ $\alpha_{54}^{1400}$.}. 

Our analysis procedure is described in Sect.\ \ref{sec:sample} and the results with a brief discussion are presented in Sect.\ \ref{sec:results}, followed  by our conclusions summarized in Sect.\ \ref{sec:conc}.
\section{The sample}\label{sec:sample}
To recall, VCSS has been carried out at the central frequency of 340~MHz with a 33.6 MHz bandwidth, resulting in $\sim$\,15\arcsec\,  resolution images, with an average rms noise ($\sigma$) of $\sim$\,3~\mjybeam. \citet{Peters2021} have provided with the VCSS epoch 1 data release a bright source catalogue (VCSS\footnote{See, \url{https://cirada.ca/vcsscatalogue}} henceforth) comprising of 52844 sources, covering 30,000 \sqdeg\, of the sky. The completeness of this VCSS catalogue is estimated to be $\sim$\,90\% for sources with a total flux density exceeding 0.2~Jy at 340 MHz and the present analysis only considers the subset of 42890 sources above this flux density.

As mentioned above, our decametric sample of sources has been extracted from the LoLSS-PDR survey at 54 MHz. The survey data release covers a sky area of $\sim$\,740~\sqdeg, imaged  with a circular beam of 47\arcsec, reaching a median rms noise level ($\sigma$) 5~\mjybeam, and resulting in a catalogue of 25247 sources. LoLSS is the first survey of such high quality, made at a very low frequency, as it provides an unprecedented combination of sensitivity and resolution. \citet{Gasperin2021} estimate that the LoLSS-PDR catalogue is 90\% complete for sources stronger than 40 mJy. In this analysis, we have only used the sources (7746) with an integrated flux density above 200~mJy at 54 MHz. As mentioned below, both our samples only consist of sources with structural codes S and M.

Here, it may be recalled that in these recent large surveys,  identification of individual discrete radio sources as well as their structural classification have been performed using an automated procedure which employs PyBDSF \citep{pybdsf}. Accordingly, a source is classified either as `Single' (type `S', resolved or unresolved), or `Multiple' (type `M', fitted with multiple Gaussian components). For both types, integrated flux densities are available in the published databases. For the present purpose, we have selected sources having either of these two structural codes, in order to arrive at samples that should be unbiased with respect to radio structural details. Thus, our 340 MHz sample consists of 42890 S/M sources above the 200 mJy limit adopted here. Out of these, NVSS counterparts could not be found for 1272 ($\sim$\,3\%) sources (see below). Their numbers in different flux density bins are listed in Tab.\ \ref{tab:vcssnvss}. 
For our 54 MHz sample containing 7746 S/M class sources above the 200 mJy limit adopted here, NVSS counterparts could not be found for 1629 ($\sim$\,21\%) sources (see below). Their numbers in different flux density bins are listed in Tab.\ \ref{tab:lolssnvss}. 

The cross-matching of the above two samples defined at 340 MHz and 54 MHz, with the 1.4~GHz NVSS was done on the basis of positional coincidence for which we have adopted a search radius of 10\arcsec. This criterion could be more effectively applied for the LoLSS sample because the beamsize  of this 54 MHz survey (47\arcsec) is nearly identical to that of the NVSS (45\arcsec) and consequently the imaging responses to the source structure are expected to be very similar at the two frequencies. Thus, if an NVSS counterpart was not found for a source in our 54 MHz sample, we have (conservatively) assigned to its 1.4 GHz flux density an upper limit of 9 mJy which is 20 times the typical rms error of 0.45 mJy in the NVSS.
 
On the other hand, the beamwidth of the 340 MHz VCSS is $\sim$\,3 times smaller than that of the NVSS (45\arcsec). This large difference could engender substantially different imaging responses to the source structure at the two frequencies, potentially amplifying the offset between the positions of the same source measured in the two surveys. Consequently, not finding an NVSS counterpart within the 10\arcsec search radius would not necessarily imply that the source is below the NVSS detection limit (taken here to be 9 mJy at 1.4 GHz, as mentioned above). Therefore, for the VCSS sources remaining unidentified in the NVSS catalog (dubbed here as `NVSS-undetected' or just `ND') it would be unrealistic to use the 9 mJy limit for constraining $\alpha_{340}^{1400}$ and for such ND sources, $\alpha_{340}^{1400}$ has been taken here as unconstrained (see below). The numbers of ND sources in the different flux density ranges are given in Tables. \ref{tab:vcssnvss} \& \ref{tab:lolssnvss}.

\begin{table*}
\centering
\setlength{\tabcolsep}{14.0pt}
\renewcommand{\arraystretch}{1.3} 
\caption{An overview of the analysis results for the VCSS$\_$NVSS combination, for different flux density ranges/bins (column 1) whose median values (mJy) are shown in column (2). Column (3) gives the number of sources in the respective bins. The corresponding median spectral indices are listed in column (4), as explained in Sect.\ \ref{sec:results1}. In column (5) number of NVSS non-detected(ND) sources are listed as explained in Sect.\ \ref{sec:sample}.}
\begin{tabular}{lccccc}
\hline
Range/Bin &Range of S$_{\rm 340~MHz}$ & S$_{\rm 340~MHz}$(median) &  Number &  $\alpha_{340}^{1400}$(median)   &  ND \\
(0)&(1) & (2) & (3) & (4) & (5) \\
\hline
All sources          & 200.0 - 135306 mJy         &             461 & 42890 &      -0.811 $\pm$    0.003 &  1272  \\
\hline
 R1          & 200.0 - 235.7 mJy     &             218 &  3574 &     (-0.765)	-0.763	(-0.760)  &  45 \\
 R2          & 235.7 - 269.9  mJy    &             252 &  3575 &      (-0.782)	-0.778	(-0.775) &  55 \\
 R3          & 269.9 - 307.5 mJy     &             288 &  3573 &      (-0.792)	-0.787	(-0.782) &  64 \\
 R4          & 307.5 - 350.0 mJy    &             328 &  3574 &   (-0.795)	-0.791	(-0.786) &  59 \\
 R5          & 350.0 - 400.5 mJy     &             374 &  3575 &  (-0.795)	-0.789	(-0.783)&  85 \\
 R6          & 400.5 - 461.1 mJy     &             430 &  3573 &   (-0.808)	-0.801	(-0.794) &  107 \\
 R7          & 461.1 - 540.1 mJy    &             498 &  3574 & (-0.808)	-0.800	(-0.792)&  119 \\
 R8          & 540.1 - 648.3 mJy    &            590 &  3575 &  (-0.821)	-0.815	(-0.808) &  125 \\
 R9          & 648.3 - 803.2 mJy   &            716 &   3574 &  (-0.823)	-0.814	(-0.806) &  132 \\
 R10         & 803.2 - 1071.4 mJy    &            921 &  3574 &    (-0.834)	-0.825	(-0.816) &  148  \\
 R11         & 1071.4 - 1676.3 mJy    &            1299 & 3575 &    (-0.850)	-0.840	(-0.828) &  167 \\
 R12         & 1676.3 - 135306 mJy &            2547 &   3574 &      (-0.862)	-0.851	(-0.838) &  166 \\
\hline
\end{tabular}
\label{tab:vcssnvss}
\end{table*}

\begin{table*}
\centering
\setlength{\tabcolsep}{14.0pt}
\renewcommand{\arraystretch}{1.3} 
\caption{An overview of the analysis results for the LoLSS$\_$NVSS combination, for different flux density bins (column 1) whose median values (mJy) are shown in column (2). Column (3) gives the number of sources in the respective flux density bins. The corresponding median spectral indices (with median error) are listed in column (4). The number of sources whose NVSS counterparts could not be identified (ND) is given in column (5) for each bin. The last column (6) gives the number of the sources for which $\alpha_{54}^{1400}$ is $<$ -1.3.}.
\begin{tabular}{lcccccc}
\hline
Range/Bin &Range of S$_{\rm 54~MHz}$ & S$_{\rm 54~MHz}$(median) &  Number &     $\alpha_{54}^{1400}$(median) &  ND &  VSS\\
(0)&(1) & (2) & (3) & (4) & (5) & (6)\\
\hline
 All sources          & 200 - 129710 mJy    &              439 & 7746 &      -0.860 $\pm$   0.004 &  1629 & 699\\
  \hline
 R1          & 200.0 - 235.0 mJy    &              218 &  969 &      -0.840 $\pm$  0.008 &  192  & 5 \\
 R2          & 235.0 - 278.4 mJy    &              255 &  968 &      -0.852 $\pm$  0.008 &  193  & 10\\
 R3          & 278.4 - 343.7 mJy    &              307 &  968 &      -0.857 $\pm$  0.009 &  204  & 12\\
 R4          & 343.7 - 438.9 mJy    &              386 &  968 &      -0.857 $\pm$  0.009 &  210  & 18\\
 R5          & 438.9 - 593.7 mJy    &              501 &  968 &      -0.860 $\pm$  0.010 &  187  & 10\\
 R6          & 593.7 - 867.0 mJy    &              706 &  968 &      -0.872 $\pm$  0.011 &  222  & 200\\
 R7          & 867.0 - 1554.0 mJy   &             1116 &  968 &      -0.862 $\pm$  0.013 &  219  & 229\\
 R8          & 1554 -  129710 mJy   &             2507 &  969 &      -0.870 $\pm$  0.018 &  202  & 215\\
\hline
\end{tabular}
\label{tab:lolssnvss}
\end{table*}
\section{Results and Discussion}\label{sec:results}
In this section, we present the  `$\alpha_{\rm median}$ - S' relations derived for our above-defined source sample selected at 340 MHz and then also compare it with its counterpart determined here at a much lower selection frequency of 54 MHz. As mentioned above, the present investigation is focused on the flux density range  $\sim$\,0.2 $<$ S$_{\rm 340}$ $<$ 2 Jy, which is dominated by intrinsically powerful, distant radio sources of synchrotron emission, having a median redshift of $\sim$\,1 \citep{Condon1989}. For several classes of intrinsically weak radio sources, including radio-quiet quasars and starburst dominated active galaxies, dependence of radio spectral index on parameters such as Eddington luminosity and ionisation ratios (across the optical diagnostic diagrams) have been investigated recently  by \citet{Laor2019} and \citet{Zajacek2019}.

\subsection{The $\alpha$-S relation at 340 MHz}\label{sec:results1}
\begin{figure*}
\centering
\includegraphics[scale=0.33]{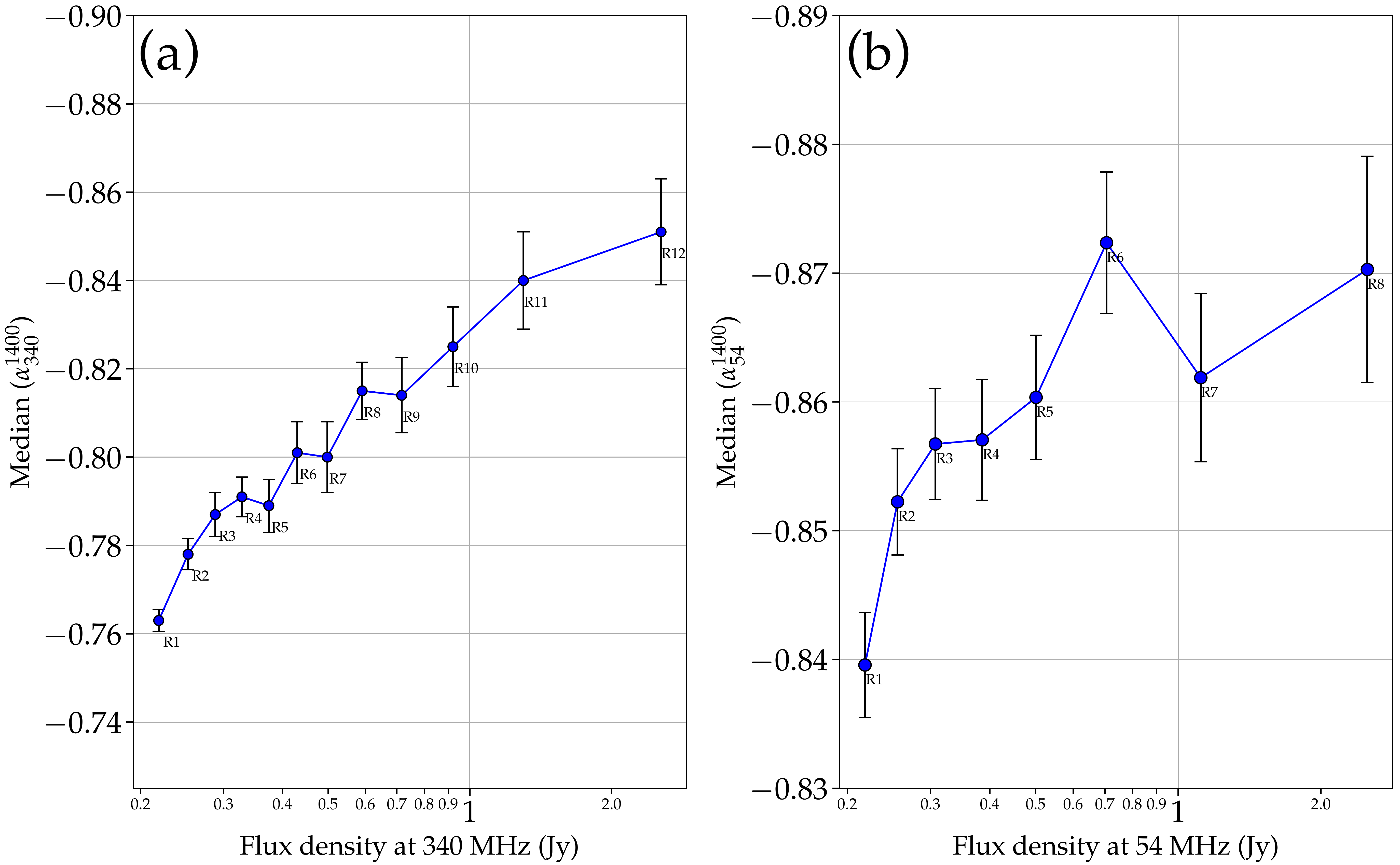} 
\caption{The above two plots show the `spectral index - flux density' relation for two different frequency samples. (a): The $\alpha_{\rm median}$- S$_{\rm 340~MHz}$ plot (Sect.\ \ref{sec:sample} $\&$ Sect.\ \ref{sec:results1}), where the data for each of the 12 flux density bins are listed in Tab.\ \ref{tab:vcssnvss}. 
(b):The $\alpha_{\rm median}$- S$_{\rm 54~MHz}$ plot (Sect.\ \ref{sec:sample} \& Sect.\ \ref{sec:result2}), where the data for each of the 8 flux density bins are listed in Tab.\ \ref{tab:lolssnvss}.
}
\label{fig:vcssslpah}
\end{figure*}

This frequency is close to the ones at which  $\alpha$-S relation has been commonly reported (Sect.\ \ref{sec:intro}) and hence the $\alpha_{\rm median}$- S$_{\rm 340~MHz}$ relation determined here can be directly compared with them. We have divided our 340~MHz VCSS sample into 12 equally populated bins of S$_{\rm 340~MHz}$ (see 
Tab.\ \ref{tab:vcssnvss} for details). Fig.\ \ref{fig:histsvcss} shows histogram of $\alpha_{340}^{1400}$ for each of the 12 flux density bins/ranges. For each histogram we have estimated 3 values of $\alpha_{340}^{1400}$(median), listed in Tab.\ \ref{tab:vcssnvss}. The middle value was determined using only the sources (97\%) for which an NVSS counterpart was found. In order to quantify the maximum impact on this estimate of $\alpha_{340}^{1400}$(median) due to the ND sources (for which $\alpha_{340}^{1400}$ is taken as unconstrained, see above),  we have considered two extreme situations about their spectral indices. In the first instance, leading to $\alpha_{340}^{1400}$(median){\it min}, we assume \textit{all} the ND sources to have $\alpha_{340}^{1400}$ well below  $\alpha_{340}^{1400}$(median), say  $\alpha_{340}^{1400}$ $<$-1.0.  In the other extreme, leading to $\alpha_{340}^{1400}$(median){\it max}, 
we assume \textit{all} the ND sources to have $\alpha_{340}^{1400}$ well above $\alpha_{340}^{1400}$(median), say $\alpha_{340}^{1400}$
$>$ -0.5. Tab.\ \ref{tab:vcssnvss} gives all the 3 estimates of $\alpha_{340}^{1400}$(median) for each flux density bin and the same are shown in Fig.\ \ref{fig:vcssslpah}(a) as the uncertainty in $\alpha_{340}^{1400}$(median) estimated for each flux bin. Despite the \textit{maximum} uncertainty of $\alpha$(median) we have taken into consideration here, the overall trend is clear and it shows a progressive flattening of $\alpha_{340}^{1400}$(median) towards lower flux densities, over the region $\sim$\,0.2 Jy $<$ S$_{\rm 340~MHz}$ $<$ $\sim$\,2 Jy. This confirms the pattern first reported in \texttt{GS82} and \texttt{SG82} and also in some (though not all) subsequent studies (Sect.\ \ref{sec:intro}). Thus, the various claims to the contrary (see Sect.\ \ref{sec:intro}) are not supported by the much more extensive and sensitive data that have now become available, after the passage of a few decades. Note, however, that the plot in Fig.\ \ref{fig:vcssslpah}a does not extend much beyond the flux density corresponding to the peak of $\alpha_{\rm median}$, somewhat limiting the scope for comparison with previous results which extended to higher flux density ranges (although at such higher flux density levels, most claims about $\alpha$-S relation are mutually consistent, anyway, see Sect.\ \ref{sec:intro}).

\begin{figure}
\centering
\includegraphics[scale=0.33]{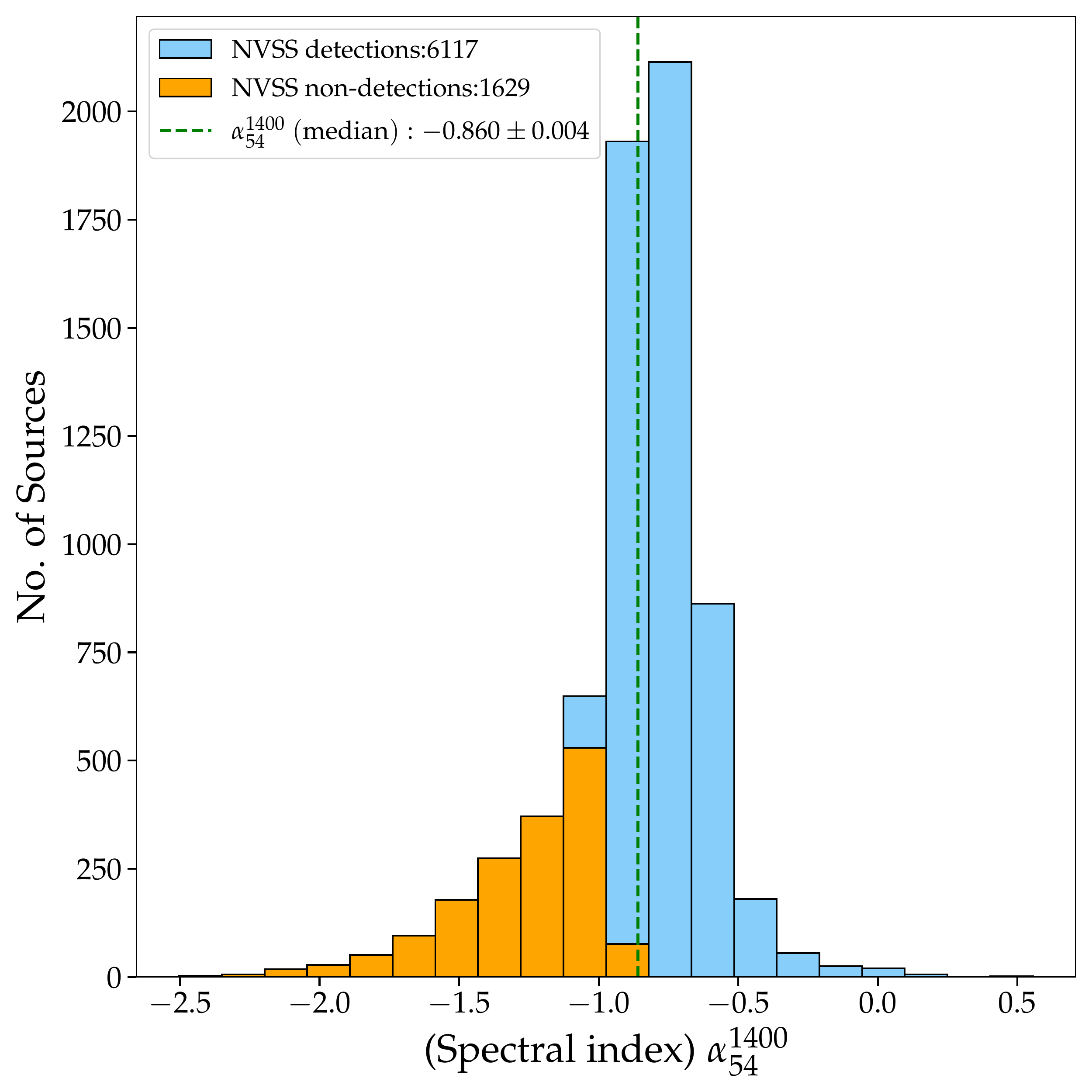} 
\caption{$\alpha_{54}^{1400}$ values of orange shaded columns are upper limits (as they represent ND sources). The value of $\alpha$(median) given inside the box is for the entire histogram.}
\label{fig:lolssalphahistfull}
\end{figure}

\subsection{The $\alpha$- S relation for sources selected at 54 MHz}\label{sec:result2}

The distribution of $\alpha_{54}^{1400}$ for our 54 MHz sample containing 7746 sources is shown in Fig.\ \ref{fig:lolssalphahistfull}. Whereas the main hump (blue shaded) is seen to drop to an insignificant level by $\alpha_{54}^{1400}$ $\sim$\,-1.3, the region at still steeper spectral indices (orange shaded) is populated by the ND sources ($\sim$\,21\%, Tab.\ \ref{tab:lolssnvss}).  This secondary yet prominent population of very-steep-spectrum (VSS) sources is likely to be associated with clusters of galaxies, as inferred from decametric radio observations conducted over the past several decades \citep[e.g.][]{BaldwinScott1973,Slingo1974,Roland1976,Bagchi1994,deGasperin2017}. Bearing this dichotomy in mind, we have proceeded to estimate $\alpha_{54}^{1400}$(median) using the entire set of sources falling in each flux density bin (Tab.\ \ref{tab:lolssnvss} ; Fig.\ \ref{fig:histlolss}), and plotted the same in
Fig.\ \ref{fig:vcssslpah}b along with median error\footnote{RMS error of the estimated median value for a histogram $= \frac{1.253\times \sigma}{\sqrt(N)}$ (here $\sigma$ is the standard deviation of the histogram and N the total number of sources in it.)}. The trend of flattening of $\alpha_{54}^{1400}$(median) with decreasing S$_{\rm 54~MHz}$ is present, though milder in comparison to that found for the 340 MHz sample (Fig.\ \ref{fig:vcssslpah}a) and confined to S$_{\rm 54~MHz}$ $<$ $\sim$\,0.5 Jy.  Conceivably, this difference could be due to the stronger presence of the VSS sources in the 54 MHz sample, as mentioned above. While this needs further investigation, 
we note that a similar onset of flattening of $\alpha_{54}^{144}$(median) near S$_{\rm 54~MHz}$ $\sim$\,0.5 Jy has been noticed in the deep LOFAR survey of a small field in Bo{\"o}tes, covering an area of $\sim$\,23.6 deg2, i.e., $\sim$\,3\% of the survey area used here (see Fig. 17 of \citealt{Williams2021}). It may also be noted that our plot shown in Fig.\ \ref{fig:vcssslpah}b extends only up to S$_{\rm 54~MHz}$ $\sim$\,3 Jy, which is roughly equivalent to S$_{\rm 340~MHz}$ $\sim$\, 0.7 Jy, i.e., well below the peak of the $\alpha_{\rm median}$ established for metre-wavelengths samples (Fig.\ \ref{fig:vcssslpah}a). Therefore, a more detailed comparison awaits future releases of LoLSS.

\section{Conclusions} \label{sec:conc}
By combining the recently released extensive high-quality database at 340 MHz (VCSS) with the NVSS at 1.4 GHz, we have revisited the spectral index - flux density relation for extragalactic radio sources selected at metre wavelengths. Here, we have focused on the flux density range below S$_{\rm 340~MHz}$ $\sim$\,1 Jy for which grossly divergent claims have been reported concerning the gradient of the $\alpha_{\rm median}$- S relation. The present work demonstrates that $\alpha_{\rm median}$ becomes progressively flatter towards decreasing flux densities below S$_{\rm 340~MHz}$ $\sim$\,1 - 2 Jy where $\alpha_{\rm median}$ has been known to attain its steepest value. This is in accord with the trend initially reported in the 1980s \citep{GS82,SG84}, also in some recent studies \citep{Tiwari2019,deGasperin2018} using the TGSS ADR1 survey at 147 MHz. 

A novel development stemming from the present study is the determination of $\alpha_{\rm median}$- S relation for extragalactic sources selected at decametre wavelengths. While broadly in agreement with the trend found for sources selected at metre wavelengths (Sect.\ \ref{sec:results1}), this relation shows a milder gradient  and appears to be less sharply peaked. A possible contributor to the difference between the two $\alpha$-S relations may be a distinctly stronger presence of very steep spectrum (VSS with $\alpha_{54}^{1400}$ $<$ -1.3) sources in the 54 MHz sample where they account for $\sim$\,9\% of the sources. Such VSS sources have long been known to be associated with clusters of galaxies and are found conspicuously in decametric radio surveys. It should be possible to verify this different trend when the second (and subsequent) releases of the LoLSS 54 MHz survey become available in the coming years. Equally promising in this context is the ongoing 14-30 MHz LOFAR Decameter Sky Survey (LoDeSS; Groeneveld et al. in prep).

A major advance in these studies would, of course, be achieved when it becomes possible to determine the $\alpha_{\rm median}$ - flux density relation using radio K-corrected values of $\alpha$ of individual sources (i.e., measured at the same frequency in the rest-frame). There has been little progress on this front since the initial attempts using metre-wavelength samples of radio sources that were comparatively small but had nearly complete redshift information \citep{GK88,Mangalam1995,Blundell1999}.

\section*{Acknowledgements}
The authors would like to thank an anonymous reviewer for helpful comments.
GK thanks Indian National Science Academy for a Senior Scientist position. GK would like to dedicate this work to the memory of Dr Hans Steppe, jointly with whom the turnover in the spectral index - flux density relation was first found.
\bibliographystyle{aa} 
\bibliography{S-alpha_20JuneFinal.bib}

\appendix
\section{Additional figures} \label{sec:appendixA}
\onecolumn
\begin{figure}
\centering
\includegraphics[scale=0.37]{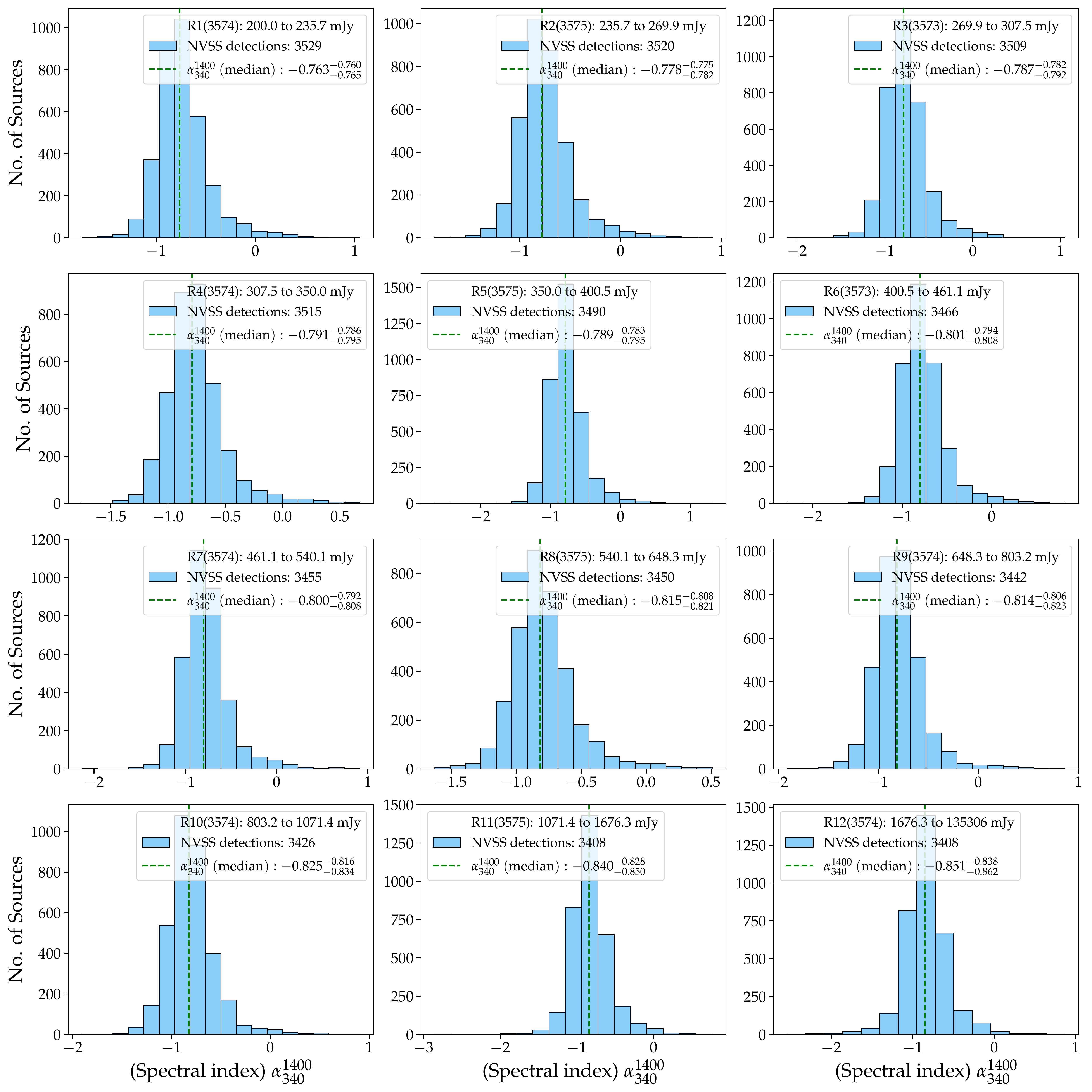}
\caption{Histogram of $\alpha_{340}^{1400}$ corresponding to only the NVSS detected sources, for each of the 12 flux-density bins (see Tab.\ \ref{tab:vcssnvss}). The 3 values of $\alpha$(median) shown inside the box for each histogram are explained in Sect.\ \ref{sec:results1}.}
\label{fig:histsvcss}
\end{figure}
\newpage

\begin{figure}
\centering
\includegraphics[scale=0.36]{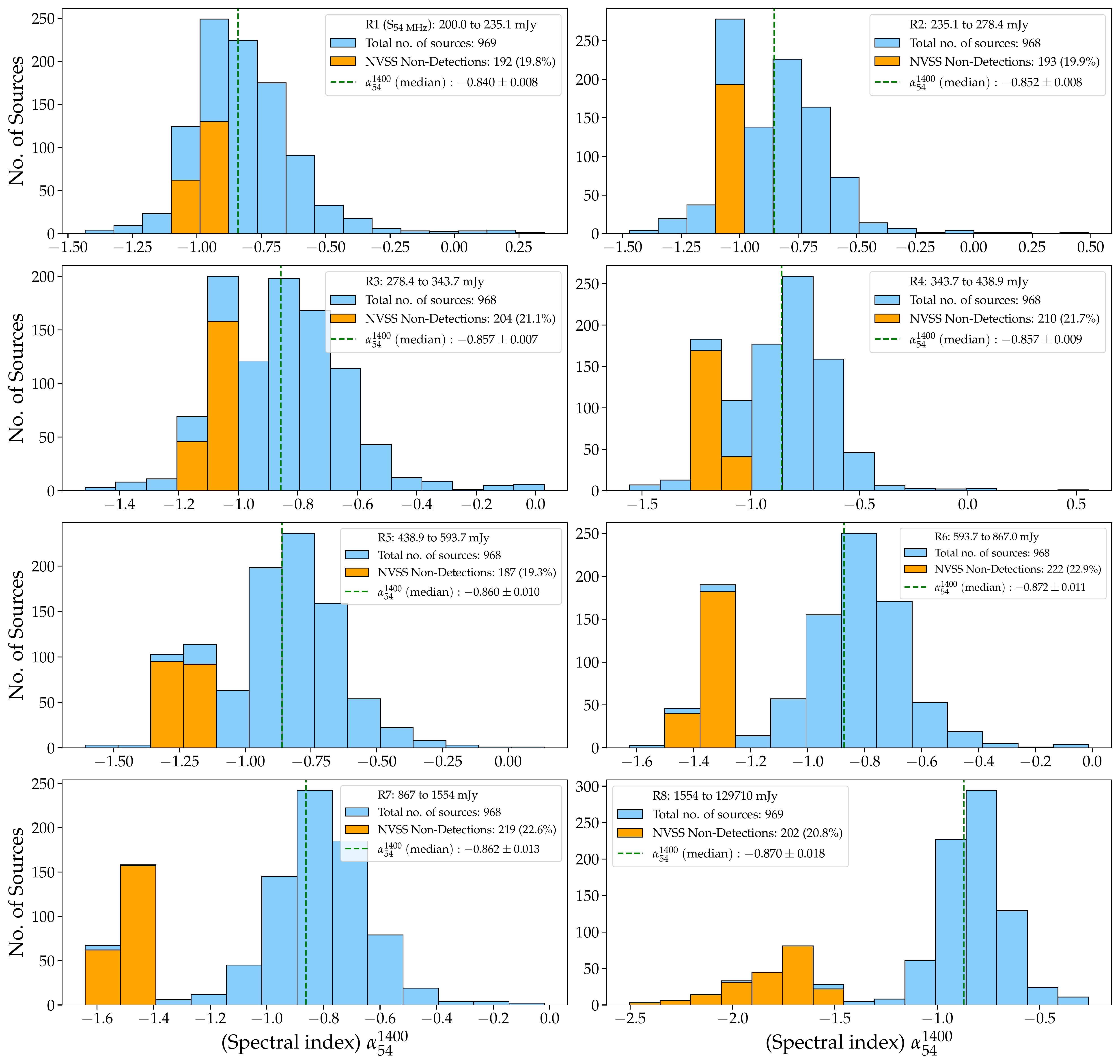}
\caption{Histogram of $\alpha_{54}^{1400}$  for each of the 8 flux-density bins. The $\alpha_{54}^{1400}$ for NVSS non-detections (ND) are shown in orange colour, and these are upper limits (Sect.\ \ref{sec:result2}). The median $\alpha_{54}^{1400}$ is indicated by green dashed line, for each histogram (see Table \ref{tab:lolssnvss}).}
\label{fig:histlolss}
\end{figure}

\end{document}